# Ultraviolet Photostability Improvement for Autofluorescence Correlation Spectroscopy on Label-Free Proteins

Aleksandr Barulin, Jérôme Wenger*

*Aix Marseille Univ, CNRS, Centrale Marseille, Institut Fresnel, 13013 Marseille, France*

* Corresponding author: jerome.wenger@fresnel.fr

**Abstract**

The poor photostability and low brightness of protein autofluorescence have been major limitations preventing the detection of label-free proteins at the single molecule level. Overcoming these issues, we report here a strategy to promote the photostability of proteins and use their natural tryptophan autofluorescence in the ultraviolet (UV) for fluorescence correlation spectroscopy (FCS). Combining enzymatic oxygen scavengers with antioxidants and triplet state quenchers greatly promotes the protein photostability, reduces the photobleaching probability and improves the net autofluorescence detection rate. Our results show that the underlying photochemical concepts initially derived for organic visible fluorescent dyes are quite general. Using this approach, we achieved UV fluorescence correlation spectroscopy on label-free streptavidin proteins containing only 24 tryptophan residues, 6.5× less than the current state-of-the-art. This strategy greatly extends the possibility to detect single label-free proteins with the versatility of single molecule fluorescence without requiring the presence of a potentially disturbing external fluorescent marker. It also opens new perspectives to improve the UV durability of organic devices.

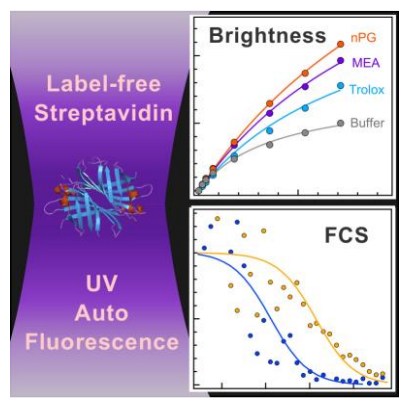

Figure for Table of Contents

Fluorescence spectroscopy and related methods such as Förster resonance energy transfer (FRET) and fluorescence correlation spectroscopy (FCS) are widely used to investigate protein structures and protein interaction dynamics.[1–3] However, the need to associate the protein of interest with an external fluorescent label (organic dye, quantum dot, nanoparticle or another protein member of the green fluorescent protein family) can turn out being a major challenge for two main reasons. First, the presence of the fluorescent marker can alter the protein structure and perturb its reactivity.[4–9] Second, the purification of the fluorescently labelled protein sample may lead to an unacceptable loss of protein material and low concentrations of the final product. An elegant approach to circumvent all the issues related with the external fluorescence labelling would be to use the intrinsic protein autofluorescence occurring in the ultraviolet (UV) range. This UV autofluorescence emission is naturally present in more than 90% of the proteins which contain at least one aromatic amino acid residues such as tryptophan, tyrosine and phenylalanine.[1,10]

Despite its immediate availability, the protein natural UV autofluorescence remains marginally used for single molecule detection and related techniques. Even for tryptophan which is the brightest aromatic amino acid, several issues contribute to challenge its detection within single proteins. First, the absorption cross-section for tryptophan is typically 10 to 50 times lower than conventional organic fluorescent dyes.[1] Second, its quantum yield is only 13% in water and is often reduced by an additional order of magnitude inside proteins due to electron and energy transfer between amino acid residues.[11,12] Third, the limited photostability of tryptophan prevents the use of high excitation powers to compensate for the low absorption cross-section and emission quantum yield.[13–16] The photostability issue is likely the most drastic limitation for single molecule fluorescence detection and FCS experiments, where the fluorescence emission rate per molecule is a crucial figure of merit. Despite these challenges, a few earlier works have pioneered the UV autofluorescence detection of single proteins using one,[13,17–19] two,[14,20] or three[15] photon excitation schemes. Due to the limited signal and photostability, only large proteins containing more than 150 tryptophan residues could be detected so far. The UV photodegradation also affects other scientific fields such as deep-UV Raman spectroscopy[59,60] or organic photovoltaics,[61–63] and has implications reaching far beyond fluorescence spectroscopy.

Looking back to the field of single molecule fluorescence detection using organic fluorescent dyes, several developments could contribute to improve the detected signal, promote the photostability and expand the technique applicability. Enzymatic oxygen scavenging systems have been introduced to remove the oxygen dissolved in the buffer solution using glucose oxidase catalase (GODCAT),[21,22] pyranose oxidase catalase (PODCAT) [23] or protocatechuate-3,4-dioxygenase / protocatechuic acid (PCD-PCA).[24] While deoxygenation improves the photostability by retarding photobleaching occurring from photo-oxidation, it also bears the negative consequence of promoting the triplet state build-up, which leads to triplet blinking and quenches the detected fluorescence signal. To avoid the negative effects of oxygen removal, various chemicals have been added to the oxygen scavenging systems in order to quench the triplet state, recover photoionized radicals and promote the photostability. These antifading agents include (i) reductants such as ascorbic acid (AA),[24,25] n propyl gallate (nPG),[24,25] Trolox,[26,27] or DABCO,[24,28] (ii) the combination of a reducing and oxidizing system (ROXS, ascorbic acid and methyl viologen) [28,29] and (iii) triplet state quenchers such as 2-mercaptoethylamine (MEA, cysteamine) [30] or β-mercaptoethanol (BME).[31] Several approaches have thus been documented for organic fluorescent dyes in the visible spectral range. However, very little is known about their application to the protein autofluorescence in the UV range, where achieving the brightest emission rate and promoting the photostability are currently the two main challenges inhibiting the detection of single label-free proteins. Besides, while photobleaching has been thoroughly investigated with rhodamines [32–35] and coumarins [36,37] dyes, the photobleaching of protein autofluorescence remains significantly less explored.[38–41]

Here, we investigate a range of strategies to promote the autofluorescence photostability of label-free proteins in the ultraviolet. The combination of enzymatic oxygen scavenging systems together with antioxidants (ascorbic acid, nPG, Trolox and DABCO) and triplet state quenchers (MEA) promotes the protein photostability, reduces the photobleaching probability and improves the autofluorescence linearity at high UV excitation rates. Our results show that the concepts initially derived for organic fluorescent dyes in the visible spectral range can be successfully applied to the intrinsic UV autofluorescence of aromatic amino-acid residues. Using an optimized approach to promote the UV photostability, we could perform UV fluorescence correlation spectroscopy on label-free streptavidin proteins containing as few as 24 tryptophan residues, 6.5× less than the standard state-of-the-art limit of detection. The systems presented here significantly extend the applicability of single label-free protein autofluorescence detection, and opens new perspectives to tune the photodynamic properties of UV emitters and increase the durability of organic photonic devices.

We first consider the case of pure tryptophan for which detailed information about photokinetic rates is avalaible,[42] and use this information to validate our approach. Figure 1a shows the basic electronic state model for tryptophan with the ground state $S_0$ and the excited state $S_1$, connected by the excitation rate $k_{ex}$, the radiative emission rate $k_r$ and the internal conversion nonradiative rate $k_{nr}$. Intersystem crossing occurs from $S_1$ to the triplet state $T_1$ with a rate constant $k_{isc}$. The triplet state is then depopulated to the ground state with a rate $k_t$. Photo-oxidation occurs from $S_1$ to the radical state $R_1$ with a rate constant $k_{ox}$. The oxidized radical can then be reduced from $R_1$ to $S_0$ with a rate $k_{red}$. Photobleaching is assumed to occur from any excited state ($S_1$, $T_1$, $R_1$) with a rate constant $k_b$ as currently the technique is not able to distinguish between photobleaching occurring from the excited singlet, triplet or radical state.[32–35] Under the steady-state assumption, the photobleaching probability $p_b$ is defined as the ratio of the number of photobleached molecules by the number of excited molecules in the $S_1$ state per unit of time. Using the rate definitions, it can be shown that $p_b = k_b/k_0$ where $k_0 = k_r + k_{nr} + k_{isc} + k_{ox}$ is the inverse of the fluorescence lifetime.[32,33] Taking into account photobleaching, the system of rate equations for the model on Fig. 1a can then be solved to express the detected average fluorescence signal F as:[32–35]

$$F = \psi \frac{\phi_f}{p_b} \left(1 - \exp\left(-\frac{4}{3} k_b S_{1eq} \tau_d\right)\right) \quad (1)$$

which includes the microscope collection efficiency $\psi$, the fluorescence quantum yield $\phi_f = k_r/k_0$, the photobleaching probability $p_b = k_b/k_0$, the steady state population of the first excited singlet state $S_{1eq}$, and the characteristic diffusion time $\tau_d$ used in FCS. The fluorescence signal dependence with the excitation rate $k_{ex}$ is contained within the population $S_{1eq}$, which can be expressed as:[43]

$$S_{1eq} = \frac{k_{ex}/k_0}{1 + \frac{k_{ex}}{k_0}\left(1 + \frac{k_{isc}}{k_t} + \frac{k_{ox}}{k_{red}}\right)} N_{tot} \quad (2)$$

where $N_{tot}$ is the total number of molecules in the detection volume. Within the model of Eqs.(1,2), the fluorescence saturation occurs when the excitation rate approaches $k_{sat} = k_0/\left(1 + \frac{k_{isc}}{k_t} + \frac{k_{ox}}{k_{red}}\right)$. Hence the buildups of the triplet state (expressed by $k_{isc}/k_t$) and of the radical state (determined by $k_{ox}/k_{red}$) play a key role in setting the transition to fluorescence saturation. Oxygen removal increases $k_{isc}/k_t$ as the triplet state is no longer quenched by dissolved oxygen, therefore it is expected that saturation occurs at a lower excitation power in the absence of oxygen in the solution. However, in addition to the triplet state contribution, the radical state also plays a major role in the occurrence of fluorescence saturation *via* the ratio $k_{ox}/k_{red}$. Determining which state (triplet or radical) is the most limiting the fluorescence brightness requires some detailed information about the photokinetic rates. We use the independent measurements performed in [42] to estimate that for tryptophan $k_{isc}/k_t$ amounts to 38 in presence of dissolved oxygen while the introduction of PODCAT oxygen scavenging system increases $k_{isc}/k_t$ to about 280 (Tab. 1). The triplet state buildup should be compared to the radical state buildup $k_{ox}/k_{red}$, which in the case of tryptophan dissolved in water amounts to 550,[42] and is thus the main limiting factor in determining the fluorescence saturation. The radical state population can be quenched by introducing reductants such as ascorbic acid (AA) to the solution, which promotes $k_{red}$ and reduces the ratio $k_{ox}/k_{red}$, down to an estimated value of 27 in presence of 10 mM ascorbic acid (Tab. 1).[42]

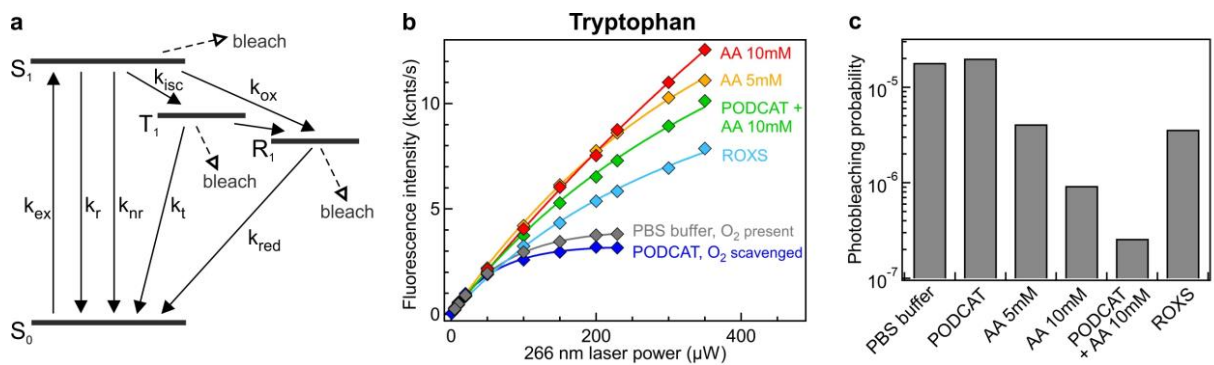

**Figure 1.** Fluorescence antifading agents to improve the photostability of pure tryptophan. (a) Electronic state model describing the tryptophan fluorescence with the ground and excited states $S_0$ and $S_1$, the triplet state $T_1$, the radical state $R_1$ and the different rate constants $k_i$. (b) Fluorescence intensity detected for 1 µM solution of tryptophan dissolved in different buffers as a function of the UV laser power. The markers are experimental data points and the lines are numerical fits based on Eq. (1). AA stands for ascorbic acid, PODCAT for pyranose oxidase catalase oxygen scavenging system, and ROXS for reducing and oxidizing system (GODCAT with 10 mM AA and 1 mM methyl viologen). The background intensity detected for oxygen scavengers and the different agents has been subtracted for each curve. (c) Tryptophan photobleaching probability $p_b$ deduced from the fits in (b) for the different antifading conditions.

**Table 1.** Evolution of the triplet state buildup $k_{isc}/k_t$ and the radical state buildup $k_{ox}/k_{red}$ for the different buffer conditions investigated experimentally in Fig. 1b. The ratios $k_{isc}/k_t$ and $k_{ox}/k_{red}$ are taken from the independent measurements reported in Ref.[42], and are used to estimate the relative change in the saturation power in our measurements.

| Buffer | $k_{isc}/k_t$ | $k_{ox}/k_{red}$ | Change of saturation power | Estimated saturation power (µW) | Measured saturation power (µW) |
|---|---|---|---|---|---|
| PBS | 38 | 550 | 1 | - | 280 ± 30 |
| PODCAT | 280 | 550 | 0.7 | 200 | 200 ± 30 |
| AA 5mM | 38 | 46 | 6.9 | 1940 | 1900 ± 200 |
| AA 10mM | 38 | 27 | 8.9 | 2500 | 2500 ± 300 |
| PODCAT + AA 10mM | 280 | 27 | 1.9 | 535 | 600 ± 80 |
| ROXS | 110 | 110 | 2.7 | 745 | 750 ± 100 |

Figure 1b shows the experimental data recorded for pure tryptophan dissolved in a water-based buffer containing 25 mM Hepes and 100mM NaCl at pH 7. This data (and similar data displayed on Fig. 2a-c for proteins) contains only the contribution F stemming from the tryptophan (or protein) autofluorescence. For each dataset, we subtract the noise contribution B stemming from the autofluorescence of the buffer solution itself from the total intensity $F_{PMT}$ recorded by the photomultiplier tube, so that we display only $F = F_{PMT} - B$. As we work at a constant concentration with a fixed confocal setup, F is directly proportional to the single molecule autofluorescence brightness or counts per molecule and the different graphs can be readily compared (F is the product of the counts per molecule times the number of molecules in the confocal detection volume $N_{mol}$).

The experimental data in Fig. 1b clearly follow the trend expected from the knowledge of the tryptophan photokinetic rates (Tab. 1): adding AA to 5 or 10 mM concentration greatly delays the occurrence of fluorescence saturation, improves the signal linearity for high excitation powers and allows reaching higher fluorescence signals. For pure tryptophan, only removing the dissolved oxygen from the solution actually degrades the fluorescence brightness and promotes saturation, as the triplet state buildup is strengthened in this case. This directly shows that for pure tryptophan the radical state buildup is the main factor determining the fluorescence saturation.

Our experimental data fit well with the model of Eq. (1), and can be used to estimate the saturation power and the photobleaching probability. The saturation power values correspond remarkably well to the data reported independently in ref.[42] for all the different buffer conditions (Tab. 1). This validates our approach combining the novel UV autofluorescence confocal microscope and the analysis based on Eqs. (1,2). For tryptophan dissolved in PBS, the photobleaching probability $p_b$ is estimated around 2e-5. Oxygen removal alone does not significantly change this value (Fig. 1c), as singlet oxygen may not be the only cause of photobleaching and as increasing the triplet state buildup may further promote the photobleaching probability by extending the residence time in the triplet state. Adding AA quenches the radical state and also reduces the photobleaching probability. Combining AA with oxygen depletion further reduces the photobleaching probability up to two orders of magnitude by emptying the oxidized state and removing single oxygen (Fig. 1c). In the case of ROXS however, the supplementary addition of methyl viologen as oxidizing agent [29] counterbalances the effects obtained with AA and PODCAT so the net photobleaching probability is reduced by one order of magnitude as compared to the PBS case, but is not as low as for the PODCAT + AA combination alone. Tryptophan's radical state appears to be cationic, and can be quenched by AA donating an electron.[42] However, the presence of methyl viologen tends to promote the radical state by accepting an electron.[29]

Having validated our approach on pure tryptophan, we now turn to protein samples. The case of protein autofluorescence is more complicated than for tryptophan alone, as in a protein the presence of the other amino acids affects the tryptophan emission.[1] In addition to the photophysics internal to tryptophan, the nearby amino acids can quench the tryptophan excited singlet state by proton transfer (lysine and tyrosine), by electron transfer (cysteine, histidine, glutamine, asparagine, glutamic acid, and aspartic acid), or by solvent quenching (depending on the exposure of tryptophan residues to water molecules).[11,12] Energy transfer between multiple tryptophan may also occur.[44] The relative influence of each of these phenomena depends on the protein structure, from the primary sequence of amino acids to the three-dimensional arrangement of polypeptide chains. As a consequence, the influence of oxygen removal or the addition of antifading agents is expected to be different for each protein, and there is no *a priori* unique antifading buffer solution that would work for all protein samples. The situation is quite similar to the organic fluorescent dyes in the visible spectral range, where the fluorescence antifades conditions need to be optimized for each different fluorophore.[24,26,28]

To provide a better understanding, we investigate here three different protein samples and explore for each of them a broad range of conditions using enzymatic oxygen scavenging systems, antioxidants and triplet state quenchers (Fig. 2). The selected proteins are streptavidin from *Streptomyces avidinii*, penicillin amidase from *Escherichia coli*, and β-galactosidase from *Escherichia coli*. Each protein contains several tryptophan amino acids to provide a higher autofluorescence signal and compensate for the intramolecular quenching induced by the other amino acids. Streptavidin has a tetrameric structure containing a total of 24 tryptophans. This protein is largely used in biochemistry owing to its high affinity for biotin. So far, it is the smallest label-free protein ever explored for confocal UV FCS. Penicillin amidase is also a multi-tryptophan protein, with a total of 29 Trp residues for which negligible inter-tryptophan energy transfer quenching was reported.[45] It appears therefore as a good candidate for a stable UV autofluorescence emission. Lastly, β-galactosidase is the largest protein investigated here, with a tetrameric structure containing 156 Trp residues. It is the first protein for which label-free UV FCS with one-photon excitation scheme was reported,[13] and it still constitutes today the current state-of-the-art.[19,46] As enzymatic scavenging system, we select PODCAT as it does not experience the significant pH drop over the course of the experiment seen for its GODCAT counterpart.[23] We also found that the protocatechuic acid (PCA) used as substrate in the PCD-PCA enzymatic scavenging system produces a large autofluorescence background in the UV which prevents the use of PCD-PCA for UV applications. Lastly as antifading agents, we explore a broad range of compounds which have been used for experiment on organic fluorescent dyes.[24–30] These agents involve 10 mM ascorbic acid (AA), 100 µM n propyl gallate (nPG), 1 mM Trolox, 10 mM DABCO, and 10 mM cysteamine (MEA).

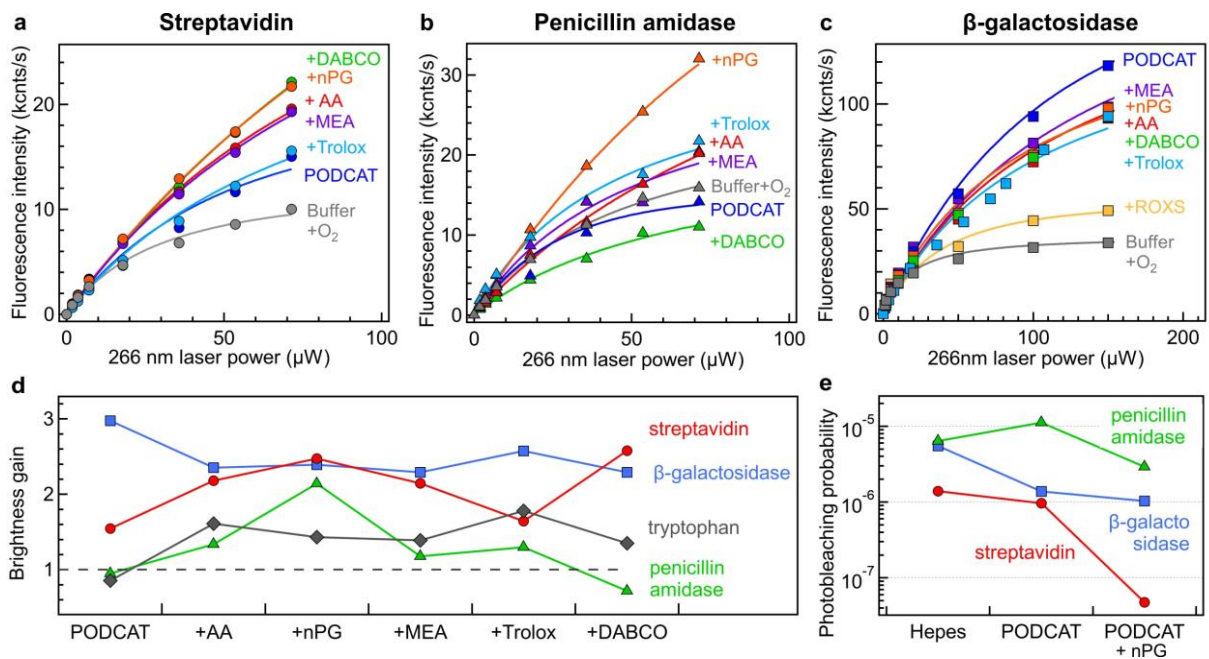

**Figure 2.** Improving the photostability of protein autofluorescence in the UV. (a-c) Fluorescence intensity detected for label-free streptavidin (a), penicillin amidase (b) and β-galactosidase (c) at 1 µM concentration in the presence of the PODCAT oxygen scavenger system plus different antifading agents. The background intensity detected for PODCAT and the different agents has been subtracted for each curve so that only the autofluorescence signal stemming from the protein is analyzed here. (d) Fluorescence brightness gains brought by the antifading agents added to the PODCAT system. The gain is expressed relative to the standard Hepes or PBS buffer solution for 100 µW excitation power at 266 nm. (e) Photobleaching probability $p_b$ deduced from the fits in (a-c) for the three different proteins with and without the addition of the PODCAT system and 100 µM nPG.

Figure 2a-c show the net detected fluorescence intensity recorded for increasing UV excitation powers for each label-free protein. For each trace, the background signal corresponding to the same buffer condition (oxygen scavenger and antifade) has been subtracted to present only the signal contribution from the protein. The total fluorescence signal increases from streptavidin to β-galactosidase as the number of Trp residues grows. However, the gain is not purely linear with the number of Trp residues as a consequence of intra-molecular autofluorescence quenching by the nearby amino-acids (Supporting Information Fig. S1). The presence of the enzymatic oxygen scavenging system PODCAT improves significantly the signal for streptavidin and β-galactosidase (Fig. 2d), it also promotes the fluorescence linearity with the excitation power and reduces the photobleaching probability (Fig. 2e). On the contrary, penicillin amidase does not follow this trend and behaves more like pure tryptophan, where the sole removal of dissolved oxygen marginally influences the autofluorescence emission. The solution to further improve the photostability is to associate the oxygen scavenger with a reductant (AA, nPG, Trolox, DABCO). The combination of both oxygen removal and radical state depletion then promotes the net signal at high excitation rates (Fig. 2d) and reduces photobleaching (Fig. 2e). For the triplet state quencher MEA, we obtain similar observations, which may indicate that it is the reducing ability of MEA which plays a role here. We also cannot exclude that the other antioxidant compounds may contribute to quench the triplet state. Importantly, we observe a gain for all the three different proteins and the different antifading agents, the only exception (out of 15 cases tested) being that DABCO appears to quench the emission of penicillin amidase even at moderate excitation powers. Out of the different compounds, nPG associated with PODCAT appears to provide the best all-around solution yielding good performance for streptavidin, penicillin amidase and β-galactosidase.

In the absence of oxygen removal, the antifading agents help to promote the photostability of streptavidin (Supporting Information Fig. S2). The best solution for the streptavidin case is to associate the reduction ability of Trolox with the triplet state quenching ability of MEA. However for penicillin amidase and β-galactosidase, no effect of the antifading agents could be detected in the presence of dissolved oxygen. For these proteins, it appears that oxygen must first be removed from the solution. Altogether, these results importantly demonstrate that the fluorescence antifading agents combined with oxygen scavengers can be used successfully to promote the UV photostability and improve the autofluorescence emission of label-free proteins.

Now that we have determined for each protein sample the best buffer configuration to promote the autofluorescence photostability and increase the autofluorescence brightness, we can use this approach to perform UV (auto)fluorescence correlation spectroscopy and characterize single label-free proteins. In FCS, the quality of the data and the signal to noise ratio are primarily determined by the fluorescence brightness per molecule.[47–50] This highlights the need for an optimization of the fluorescence emission rate and photostability, as in FCS a low signal cannot be compensated by a higher fluorophore concentration.[25] Figure 3a-c shows the FCS correlation traces for the three different proteins in the conditions yielding the optimum signal to background. The corresponding fluorescence time traces are displayed in the Supporting Information Fig. S3 for each protein and each buffer alone. We point out that in the standard hepes or PBS buffer, the FCS signal to noise is strongly degraded (Supporting Information Fig. S4) preventing the direct acquisition of FCS data on label-free proteins if no special care is taken. We have also checked that no FCS correlation is obtained in the absence of the protein sample (Supporting Information Fig. S5).[46] From the FCS data analysis (see details in the methods section below), the diffusion time $\tau_d$ (Fig 3d) and the number of proteins $N_{mol}$ in the confocal detection volume (Fig. 3e) are quantified. Ideally, other fast processes inducing a change in the autofluorescence intensity could be detected by FCS, such as triplet state blinking [51], photoinduced electron transfer (PET) [52,53] or intra-chain diffusion.[54] However, the uncorrelated blinking from the collection of a large number of Trp emitters and the limited FCS signal to noise ratio at lag times below 10 μs currently prevent any clear observation of any of these fast dynamics phenomena.

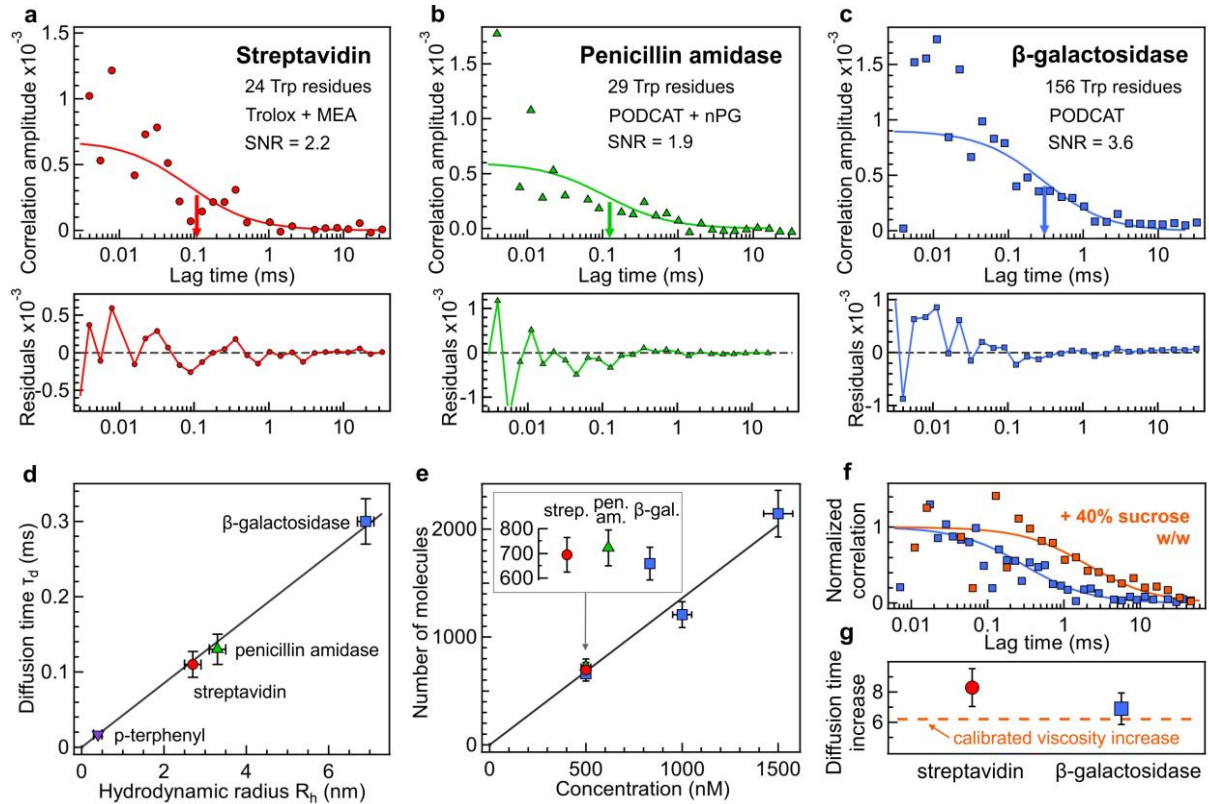

**Figure 3.** Fluorescence correlation spectroscopy experiments performed on label-free proteins in the UV range. (a-c) FCS correlation functions recorded for streptavidin (a), penicillin amidase (b) and β-galactosidase (c) at 500 nM concentration. The lines are numerical fits based on the diffusion model in Eq. (3), the arrows indicate the diffusion time $\tau_d$. The lower traces display the fit residuals. The signal to background ratio defined as SNR = F/B = ($F_{PMT}$ – B)/B is displayed on each graph. For (a) the buffer contains 1 mM Trolox and 10 mM MEA. For (b), we use the PODCAT system with 100 μM nPG while for (c) only POCDAT is used. (d) Diffusion time measured by UV FCS as a function of the molecular hydrodynamic radius. The line is computed from Stokes Einstein equation using a laser beam radius at $1/e^2$ of 195 nm (see text for details). (e) Number of proteins detected by UV FCS as a function of the sample concentration. The line corresponds to a confocal volume of 2.2 fL. (f) FCS correlation function for β-galactosidase without and with 40% w/w sucrose to increase the viscosity. (g) Increase in the diffusion time recorded by UV FCS due to the addition of 40% w/w sucrose to the solution. The orange dashed line corresponds to the viscosity increase independently measured in Ref.[55]

Several tests can be made to assess the validity of our FCS results. First, we plot the FCS diffusion time measured for each protein as a function of its calibrated hydrodynamic radius (Fig. 3d, the data for the UV fluorescent dye p-terphenyl is taken from our previous work Ref.[46] ). For streptavidin and β-galactosidase, we use the hydrodynamic radii independently measured in Ref.[56,57] and in Ref.[13] respectively. For penicillin amidase, we estimate the hydrodynamic radius from the total number of amino acid residues in the polypeptide chain, following the approach developed in Ref.[58] The FCS diffusion time scales as $\tau_d = w^2/4D$, where $w$ is the laser beam radius at 1/e² at the microscope focus and $D = k_B T/(6 \pi \eta R_h)$ is the protein translational diffusion coefficient, which is inversely proportional to the hydrodynamic radius $R_h$ and the solvent viscosity $\eta$ according to Stokes-Einstein's law. Thus the diffusion time $\tau_d$ scales linearly with the hydrodynamic radius $R_h$, and our experimental data nicely reproduces this trend. From the slope of the line in Fig. 3d we get a laser beam radius $w$ of 195 nm, which corresponds well to the $1/\sqrt{2 \ln 2} * \lambda/2NA =$ 190 nm estimate of the spot 1/e² beam radius for a $\lambda =$ 266 nm laser beam focused by a $NA = 0.6$ objective.[43]

The second set of validity tests checks that for the three different proteins at the same concentration, comparable number of molecules are measured by FCS (Fig. 3e insert). We also check by performing a series of dilutions that the number of molecules measured depends linearly with the concentration (Fig. 3e and Supporting Information Fig. S6). The third and last test goes with the addition of 40% w/w sucrose to the solution to change the viscosity $\eta$ in a calibrated manner.[55] The viscosity increase directly impacts the FCS diffusion time, and our FCS data for both streptavidin and β-galactosidase quantitatively reproduces the expected viscosity increase within experimental uncertainties (Fig. 3f,g and Supporting Information Fig. S7). It should be noted that the evolution of the viscosity with the sucrose content is highly nonlinear, and a 1% change in the sucrose amount can modify the viscosity by 10%. Altogether, the different characterizations reported in Fig. 3d-g validate the UV FCS approach and establish our ability to analyze label-free proteins. As the microscope is fully time-resolved, the same setup can also be used to provide fluorescence lifetime information (Supporting Information Fig. S8).

To conclude, the presumed poor photostability and low brightness of protein UV autofluorescence have been limiting factors preventing the direct label-free detection of proteins in the ultraviolet. The limited UV photostability has consequences affecting other scientific domains such as deep-UV surface-enhanced Raman spectroscopy[59,60] or organic photovoltaics.[61–63] Here we show that combining of enzymatic oxygen scavenging systems with chemical antifading agents can greatly promote the protein photostability, reduce the photobleaching probability and improve the net autofluorescence detection rate per molecule. These strategies, inspired by the developments for organic visible fluorescent dyes,[24–30] are now for the first time applied to the intrinsic UV autofluorescence of natural proteins, showing the their underlying photochemical concepts are quite general. The reductant nPG associated with PODCAT oxygen scavengers provides a good performance for the different proteins, while ascorbic acid, Trolox and MEA offer efficient alternatives. Using this photostability optimization, we could record FCS data on label-free proteins to measure their local concentration, diffusion coefficient and hydrodynamic radius without adding any external fluorescent marker. While previous works relied on large proteins containing more than 150 Trp residues,[13–15,17–20] here we could detect label-free streptavidin proteins containing as few as 24 Trp residues. This approach greatly extends the applicability of single label-free protein autofluorescence detection and improves our understanding of the photochemical degradation in the UV. It can be combined with nanophotonics and plasmonics [46,64–67] as well as with reverse intersystem crossing [68] to further enhance the fluorescence brightness and promote photostability. Our results open new perspectives for different scientific communities. For single molecule biophysics, a wide range of FCS studies and related techniques can now be performed on a large library of proteins,[69,70] not being restricted anymore to specific proteins featuring several hundreds of tryptophan residues. For physical chemistry, further experiments and modelling can build on our results to explore new strategies to control the photokinetics of indole-based emitters. For organic optoelectronics, our results suggest new strategies to increase the UV durability.

*Tryptophan and protein samples.* All the chemicals are purchased from Sigma Aldrich if not stated otherwise. L-tryptophan is dissolved in a Hepes buffer (Hepes 25 mM, NaCl 100mM, pH 7) at pH 7 or PBS, if indicated. The proteins used in our UV molecular autofluorescence studies are β-galactosidase from *Escherichia coli* (M=465 kDa, 156 tryptophan residues), streptavidin from *Streptomyces avidinii* (M=52.8 kDa, 24 tryptophan residues) and penicillin amidase from *Escherichia coli* (M=86 kDa, 29 tryptophan residues). The protein concentrations are assessed by a spectrophotometer (Tecan, Spark 10M). The stock solutions of protein molecules are stored at -20 °C. Before the measurements, the stock solution is slowly defrosted at 4 °C, then at the room temperature (20 °C) and diluted to concentrations indicated previously. For single molecule analyses, sodium hydroxide and 0.5% v/v Tween20 are added in the final protein mixture in order to prevent aggregate formation. The solution pH lies in proximity of 8 in those cases. Concentrations of streptavidin, penicillin amidase and β-galactosidase equal 500 nM for single molecule investigation by UV FCS. In addition, we study autocorrelation functions for β-galactosidase at several concentrations. For the viscosity increase experiment, 40 w/w% sucrose (Analytical reagent grade, Fisher Scientific) is introduced in the buffer solution.

*Oxygen scavengers, antioxidants and antifades.* D-glucose (≥99.5%), pyranose oxidase (from Coriolus sp., expressed in *E.coli*, M=270 kDa), glucose oxidase (from Aspergillus niger, M=160 kDa) and catalase (from bovine liver, M=250 kDa) are exploited in this work. PODCAT oxygen scavenger mixtures contain 10 wt% of D-glucose, 0.83 µM of catalase and 1.5 µM of pyranose oxidase. GODCAT solution contain 10 wt% of D-glucose, 0.83 µM of catalase and 0.3 µM of glucose oxidase. We found similar results for these PODCAT and GODCAT systems. The lower activity of pyranose oxidase (as compared to glucose oxidase) is compensated by a higher concentration of this enzyme. Stock solutions of oxygen scavenger enzymes are stored in the hepes buffer at -20 °C, while a D-glucose stock solution is stored in Milli-Q water at 4 °C.

Ascorbic acid (AA) is dissolved in Milli-Q deionized water at 300 mM concentration, and we add 1M NaOH to bring it to pH7. Then, the stock solution is diluted to 10 mM for all corresponding experiments, if not indicated otherwise. Mercaptoethylamine (MEA) is dissolved at 1 M concentration, adjusting its pH to 7.5 using glacial acetic acid. The MEA solution is then diluted to reach a 10 mM concentration for our experiments. 1,4-Diazabicyclo[2.2.2]octane (DABCO) is dissolved in Milli-Q deionized water at 1 M concentration, and pH is adjusted at 7.5 by adding HCl in the solution. DABCO is diluted down to 10 mM for all corresponding experiments. *n*-Propyl gallate (nPG) stock solution is prepared at 10 mM concentration in 1 v/v% absolute ethanol/water. nPG starts promoting photostability at lower concentrations, therefore it is diluted to 100 µM. For creating ROXS-conditions, we used 1 mM solution of methyl viologen (MV) and 10 mM ascorbic acid solution in the final mixture with oxygen scavengers. Trolox is dissolved in DMSO at 100 mM concentration and is stored under aluminum foil no longer than 48 hours in order to prevent Trolox-quinone accumulation in the solution. Trolox-quinone is supposed to be present during the experiment in the confocal volume due to its UV-induced formation under the 266 nm laser illumination. The final Trolox concentration is 1 mM in the investigated solutions.

*UV FCS microscope.* Ensemble fluorescence measurements and FCS experiments on individual proteins are performed using a home-built UV confocal microscope setup. The laser source is a LDH-P-FA-266 laser emitting at 266 nm (Picoquant, 70 ps pulse duration, 80 MHz repetition rate). The laser beam is spatially filtered with a 50 µm pinhole and spectrally filtered by a short pass filter (FF01-311/SP-25, Semrock). FCS experiments on β-galactosidase are performed at 150 µW laser power (measured at the microscope entrance port before the dichroic mirror), while the experiments on penicillin amidase and streptavidin are taken at 300 µW in order to compensate the lower autofluorescence signal.

The microscope uses a dichroic mirror (FF310-Di01-25-D, Semrock) to reflect the laser beam and transmit the autofluorescence signal. A Zeiss Ultrafluar 40x, 0.6 NA glycerol immersion objective focuses the UV laser beam inside a protein solution through a 0.15 µm quartz coverslip creating a diffraction limited confocal volume. The fluorescence light is collected back through the same objective and transmitted through the dichroic mirror, a long pass filter (FF01-300/LP-25, Semrock) and an emission band pass filter (FF01-375/110-25, Semrock). An air-spaced achromatic doublet with 200 mm focal length (ACA254-200-UV, Thorlabs) is used as a tube lens of the microscope. The detection channel is equipped with a 80 µm pinhole for spatial filtering of the fluorescence light. The detection of the autofluorescence light in the 310-410 spectral range is performed by a photomultiplier tube (PMA 175, Picoquant) whose output is connected to a time correlated single photon counting module (Picoharp 300, Picoquant) in a time tagged time resolved mode (TTTR). The analysis of the autofluorescence time traces is carried out with SymPhoTime 64 software (Picoquant). FCS correlation functions are extracted from time traces integrated for 5 min. The detector afterpulsing is negligible for correlation lag times longer than 1 µs.

*FCS analysis.* Fitting of correlation functions is performed with a standard Brownian diffusion model: [69,70]

$$G(t) = \frac{1}{N_{mol}} \left(1 - \frac{B}{F_{PMT}}\right)^2 \frac{1}{(1+\frac{\tau}{\tau_d})\cdot(1+\frac{\tau}{\kappa^2 \tau_d})^{0.5}} \qquad (3)$$

$B/F_{PMT}$ is the ratio of background counts B to the total detected fluorescence signal $F_{PMT}$ = F+B (where F represents the autofluorescence signal stemming only from the proteins), $N_{mol}$ is the number of molecules in the detection volume, $\tau_d$ represents the average diffusion time of the molecules through the detection volume. The aspect ratio of confocal volume κ is kept at a constant value of 8 for all the fits. This model does not consider triplet state blinking [51] nor other fast dynamics processes such as photoinduced electron transfer (PET) [52,53] or intra-chain diffusion[54], as the large number of Trp emitters and the limited signal to noise ratio in FCS at short lag times below 10 µs prevent any clear observation of these phenomena. The parameters extracted from the FCS correlation function are $N_{mol}$ and $\tau_d$, which provide information on concentration of protein molecules and diffusivity, respectively.

**Notes**

The authors declare no competing financial interest.

**Acknowledgments**

The authors thank Satyajit Patra for stimulating discussions. This project has received funding from the European Research Council (ERC) under the European Union's Horizon 2020 research and innovation programme (grant agreement No 723241).

**References**

(1) Lakowicz, J. R. *Principles of Fluorescence Spectroscopy*, 3rd ed.; Springer US, 2006.
(2) Weiss, S. Fluorescence Spectroscopy of Single Biomolecules. *Science* **1999**, *283*, 1676–1683.
(3) Kim, S. A.; Heinze, K. G.; Schwille, P. Fluorescence Correlation Spectroscopy in Living Cells. *Nat. Methods* **2007**, *4*, 963–973.
(4) Kudryashov, D. S.; Phillips, M.; Reisler, E. Formation and Destabilization of Actin Filaments with Tetramethylrhodamine-Modified Actin. *Biophys. J.* **2004**, *87*, 1136–1145.
(5) Cabantous, S.; Terwilliger, T. C.; Waldo, G. S. Protein Tagging and Detection with Engineered Self-Assembling Fragments of Green Fluorescent Protein. *Nat. Biotechnol.* **2005**, *23*, 102-107.
(6) Zanetti-Domingues, L. C.; Tynan, C. J.; Rolfe, D. J.; Clarke, D. T.; Martin-Fernandez, M. Hydrophobic Fluorescent Probes Introduce Artifacts into Single Molecule Tracking Experiments Due to Non-Specific Binding. *PLOS ONE* **2013**, *8*, e74200.
(7) Hughes, L. D.; Rawle, R. J.; Boxer, S. G. Choose Your Label Wisely: Water-Soluble Fluorophores Often Interact with Lipid Bilayers. *PLOS ONE* **2014**, *9*, e87649.
(8) Dietz, M. S.; Wehrheim, S. S.; Harwardt, M.-L. I. E.; Niemann, H. H.; Heilemann, M. Competitive Binding Study Revealing the Influence of Fluorophore Labels on Biomolecular Interactions. *Nano Lett.* **2019**, *19*, 8245–8249.
(9) Patra, S.; Baibakov, M.; Claude, J.-B.; Wenger, J. Surface Passivation of Zero-Mode Waveguide Nanostructures: Benchmarking Protocols and Fluorescent Labels. *ArXiv200104718 Phys.* **2020**.
(10) Kumamoto, Y.; Taguchi, A.; Kawata, S. Deep-Ultraviolet Biomolecular Imaging and Analysis. *Adv. Opt. Mater.* **2019**, *7*, 1801099.
(11) Chen, Y.; Barkley, M. D. Toward Understanding Tryptophan Fluorescence in Proteins. *Biochemistry* **1998**, *37*, 9976–9982.
(12) Adams, P. D.; Chen, Y.; Ma, K.; Zagorski, M. G.; Sönnichsen, F. D.; McLaughlin, M. L.; Barkley, M. D. Intramolecular Quenching of Tryptophan Fluorescence by the Peptide Bond in Cyclic Hexapeptides. *J. Am. Chem. Soc.* **2002**, *124*, 9278–9286.
(13) Li, Q.; Seeger, S. Label-Free Detection of Single Protein Molecules Using Deep UV Fluorescence Lifetime Microscopy. *Anal. Chem.* **2006**, *78*, 2732–2737.
(14) Lippitz, M.; Erker, W.; Decker, H.; Holde, K. E. van; Basché, T. Two-Photon Excitation Microscopy of Tryptophan-Containing Proteins. *Proc. Natl. Acad. Sci.* **2002**, *99*, 2772–2777.
(15) Ranjit, S.; Dvornikov, A.; Holland, D. A.; Reinhart, G. D.; Jameson, D. M.; Gratton, E. Application of Three-Photon Excitation FCS to the Study of Protein Oligomerization. *J. Phys. Chem. B* **2014**, *118*, 14627–14631.
(16) Sanabia, J. E.; Goldner, L. S.; Lacaze, P.-A.; Hawkins, M. E. On the Feasibility of Single-Molecule Detection of the Guanosine-Analogue 3-MI. *J. Phys. Chem. B* **2004**, *108*, 15293–15300.
(17) Wennmalm, S.; Blom, H.; Wallerman, L.; Rigler, R. UV-Fluorescence Correlation Spectroscopy of 2-Aminopurine. *Biol. Chem.* **2001**, *382*, 393–397.
(18) Li, Q.; Ruckstuhl, T.; Seeger, S. Deep-UV Laser-Based Fluorescence Lifetime Imaging Microscopy of Single Molecules. *J. Phys. Chem. B* **2004**, *108*, 8324–8329.
(19) Li, Q.; Seeger, S. Label-Free Detection of Protein Interactions Using Deep UV Fluorescence Lifetime Microscopy. *Anal. Biochem.* **2007**, *367*, 104–110.

(20) Sahoo, B.; Balaji, J.; Nag, S.; Kaushalya, S. K.; Maiti, S. Protein Aggregation Probed by Two-Photon Fluorescence Correlation Spectroscopy of Native Tryptophan. *J. Chem. Phys.* **2008**, *129*, 075103.

(21) Benesch, R. E.; Benesch, R. Enzymatic Removal of Oxygen for Polarography and Related Methods. *Science* **1953**, *118*, 447–448.

(22) Blanchard, S. C.; Gonzalez, R. L.; Kim, H. D.; Chu, S.; Puglisi, J. D. TRNA Selection and Kinetic Proofreading in Translation. *Nat. Struct. Mol. Biol.* **2004**, *11*, 1008–1014.

(23) Swoboda, M.; Henig, J.; Cheng, H.-M.; Brugger, D.; Haltrich, D.; Plumeré, N.; Schlierf, M. Enzymatic Oxygen Scavenging for Photostability without PH Drop in Single-Molecule Experiments. *ACS Nano* **2012**, *6*, 6364–6369.

(24) Aitken, C. E.; Marshall, R. A.; Puglisi, J. D. An Oxygen Scavenging System for Improvement of Dye Stability in Single-Molecule Fluorescence Experiments. *Biophys. J.* **2008**, *94*, 1826–1835.

(25) Widengren, J.; Chmyrov, A.; Eggeling, C.; Löfdahl, P.-Å.; Seidel, C. A. M. Strategies to Improve Photostabilities in Ultrasensitive Fluorescence Spectroscopy. *J. Phys. Chem. A* **2007**, *111*, 429–440.

(26) Rasnik, I.; McKinney, S. A.; Ha, T. Nonblinking and Long-Lasting Single-Molecule Fluorescence Imaging. *Nat. Methods* **2006**, *3*, 891–893.

(27) Cordes, T.; Vogelsang, J.; Tinnefeld, P. On the Mechanism of Trolox as Antiblinking and Antibleaching Reagent. *J. Am. Chem. Soc.* **2009**, *131*, 5018–5019.

(28) Cordes, T.; Maiser, A.; Steinhauer, C.; Schermelleh, L.; Tinnefeld, P. Mechanisms and Advancement of Antifading Agents for Fluorescence Microscopy and Single-Molecule Spectroscopy. *Phys. Chem. Chem. Phys.* **2011**, *13*, 6699–6709.

(29) Vogelsang, J.; Kasper, R.; Steinhauer, C.; Person, B.; Heilemann, M.; Sauer, M.; Tinnefeld, P. A Reducing and Oxidizing System Minimizes Photobleaching and Blinking of Fluorescent Dyes. *Angew. Chem. Int. Ed.* **2008**, *47*, 5465–5469.

(30) Campos, L. A.; Liu, J.; Wang, X.; Ramanathan, R.; English, D. S.; Muñoz, V. A Photoprotection Strategy for Microsecond-Resolution Single-Molecule Fluorescence Spectroscopy. *Nat. Methods* **2011**, *8*, 143–146.

(31) Kishino, A.; Yanagida, T. Force Measurements by Micromanipulation of a Single Actin Filament by Glass Needles. *Nature* **1988**, *334*, 74–76.

(32) Eggeling, C.; Widengren, J.; Rigler, R.; Seidel, C. A. M. Photobleaching of Fluorescent Dyes under Conditions Used for Single-Molecule Detection: Evidence of Two-Step Photolysis. *Anal. Chem.* **1998**, *70*, 2651–2659.

(33) Eggeling, C.; Volkmer, A.; Seidel, C. A. M. Molecular Photobleaching Kinetics of Rhodamine 6G by One- and Two-Photon Induced Confocal Fluorescence Microscopy. *ChemPhysChem* **2005**, *6*, 791–804.

(34) Eggeling, C.; Widengren, J.; Brand, L.; Schaffer, J.; Felekyan, S.; Seidel, C. A. M. Analysis of Photobleaching in Single-Molecule Multicolor Excitation and Förster Resonance Energy Transfer Measurements. *J. Phys. Chem. A* **2006**, *110*, 2979–2995.

(35) Eggeling, C.; Widengren, J.; Rigler, R.; Seidel, C. A. M. Photostability of Fluorescent Dyes for Single-Molecule Spectroscopy: Mechanisms and Experimental Methods for Estimating Photobleaching in Aqueous Solution. In *Applied Fluorescence in Chemistry, Biology and Medicine*; Rettig, W., Strehmel, B., Schrader, S., Seifert, H., Eds.; Springer: Berlin, Heidelberg, 1999; pp 193–240.

(36) Brand, L.; Eggeling, C.; Zander, C.; Drexhage, K. H.; Seidel, C. A. M. Single-Molecule Identification of Coumarin-120 by Time-Resolved Fluorescence Detection: Comparison of One- and Two-Photon Excitation in Solution. *J. Phys. Chem. A* **1997**, *101*, 4313–4321.

(37) Eggeling, C.; Brand, L.; Seidel, C. a. M. Laser-Induced Fluorescence of Coumarin Derivatives in Aqueous Solution: Photochemical Aspects for Single Molecule Detection. *Bioimaging* **1997**, *5*, 105–115.

(38) Nilsson, R.; Merkel, P. B.; Kearns, D. R. Unambiguous Evidence For The Participation Of Singlet Oxygen (1δ) In Photodynamic Oxidation Of Amino Acids. *Photochem. Photobiol.* **1972**, *16*, 117–124.

(39) Silvester, J. A.; Timmins, G. S.; Davies, M. J. Photodynamically Generated Bovine Serum Albumin Radicals: Evidence for Damage Transfer and Oxidation at Cysteine and Tryptophan Residues. *Free Radic. Biol. Med.* **1998**, *24*, 754–766.

(40) Matuszak, Z.; Bilska, M. A.; Reszka, K. J.; Chignell, C. F.; Bilski, P. Interaction of Singlet Molecular Oxygen with Melatonin and Related Indoles. *Photochem. Photobiol.* **2003**, *78*, 449–455.

(41) Schulze, P.; Ludwig, M.; Belder, D. Impact of Laser Excitation Intensity on Deep UV Fluorescence Detection in Microchip Electrophoresis. *Electrophoresis* **2008**, *29*, 4894–4899.

(42) Hevekerl, H.; Tornmalm, J.; Widengren, J. Fluorescence-Based Characterization of Non-Fluorescent Transient States of Tryptophan – Prospects for Protein Conformation and Interaction Studies. *Sci. Rep.* **2016**, *6*, 35052.

(43) Enderlein, J.; Zander, C. Theoretical Foundations of Single Molecule Detection in Solution. In *Single Molecule Detection in Solution*; Wiley-Blackwell, 2003; pp 21–67.

(44) Demchenko, A. P.; Gallay, J.; Vincent, M.; Apell, H.-J. Fluorescence Heterogeneity of Tryptophans in Na,K-ATPase: Evidences for Temperature-Dependent Energy Transfer. *Biophys. Chem.* **1998**, *72*, 265–283.

(45) Ercelen, S.; Kazan, D.; Erarslan, A.; Demchenko, A. P. On the Excited-State Energy Transfer between Tryptophan Residues in Proteins: The Case of Penicillin Acylase. *Biophys. Chem.* **2001**, *90*, 203–217.

(46) Barulin, A.; Claude, J.-B.; Patra, S.; Bonod, N.; Wenger, J. Deep Ultraviolet Plasmonic Enhancement of Single Protein Autofluorescence in Zero-Mode Waveguides. *Nano Lett.* **2019**, *19*, 7434–7442.

(47) Koppel, D. E. Statistical Accuracy in Fluorescence Correlation Spectroscopy. *Phys. Rev. A* **1974**, *10*, 1938–1945.

(48) Meseth, U.; Wohland, T.; Rigler, R.; Vogel, H. Resolution of Fluorescence Correlation Measurements. *Biophys. J.* **1999**, *76*, 1619–1631.

(49) Wohland, T.; Rigler, R.; Vogel, H. The Standard Deviation in Fluorescence Correlation Spectroscopy. *Biophys. J.* **2001**, *80*, 2987–2999.

(50) Wenger, J.; Gérard, D.; Aouani, H.; Rigneault, H.; Lowder, B.; Blair, S.; Devaux, E.; Ebbesen, T. W. Nanoaperture-Enhanced Signal-to-Noise Ratio in Fluorescence Correlation Spectroscopy. *Anal. Chem.* **2009**, *81*, 834–839.

(51) Widengren, J.; Rigler, R.; Mets, Ü. Triplet-State Monitoring by Fluorescence Correlation Spectroscopy. *J. Fluoresc.* **1994**, *4*, 255–258.

(52) Neuweiler, H.; Löllmann, M.; Doose, S.; Sauer, M. Dynamics of Unfolded Polypeptide Chains in Crowded Environment Studied by Fluorescence Correlation Spectroscopy. *J. Mol. Biol.* **2007**, *365*, 856–869.

(53) Doose, S.; Neuweiler, H.; Sauer, M. Fluorescence Quenching by Photoinduced Electron Transfer: A Reporter for Conformational Dynamics of Macromolecules. *ChemPhysChem* **2009**, *10*, 1389–1398.

(54) Neuweiler, H.; Johnson, C. M.; Fersht, A. R. Direct Observation of Ultrafast Folding and Denatured State Dynamics in Single Protein Molecules. *Proc. Natl. Acad. Sci.* **2009**, *106*, 18569–18574.

(55) Telis, V. R. N.; Telis-Romero, J.; Mazzotti, H. B.; Gabas, A. L. Viscosity of Aqueous Carbohydrate Solutions at Different Temperatures and Concentrations. *Int. J. Food Prop.* **2007**, *10*, 185–195.

(56) Jazani, S.; Sgouralis, I.; Shafraz, O. M.; Levitus, M.; Sivasankar, S.; Pressé, S. An Alternative Framework for Fluorescence Correlation Spectroscopy. *Nat. Commun.* **2019**, *10*, 1–10.

(57) Strömqvist, J.; Nardo, L.; Broekmans, O.; Kohn, J.; Lamperti, M.; Santamato, A.; Shalaby, M.; Sharma, G.; Di Trapani, P.; Bondani, M.; et al. Binding of Biotin to Streptavidin: A Combined

Fluorescence Correlation Spectroscopy and Time-Resolved Fluorescence Study. *Eur. Phys. J. Spec. Top.* **2011**, *199*, 181–194.

(58) Wilkins, D. K.; Grimshaw, S. B.; Receveur, V.; Dobson, C. M.; Jones, J. A.; Smith, L. J. Hydrodynamic Radii of Native and Denatured Proteins Measured by Pulse Field Gradient NMR Techniques. *Biochemistry* **1999**, *38*, 16424–16431.

(59) Jha, S. K.; Ahmed, Z.; Agio, M.; Ekinci, Y.; Löffler, J. F. Deep-UV Surface-Enhanced Resonance Raman Scattering of Adenine on Aluminum Nanoparticle Arrays. *J. Am. Chem. Soc.* **2012**, *134*, 1966–1969.

(60) Sharma, B.; Cardinal, M. F.; Ross, M. B.; Zrimsek, A. B.; Bykov, S. V.; Punihaole, D.; Asher, S. A.; Schatz, G. C.; Van Duyne, R. P. Aluminum Film-Over-Nanosphere Substrates for Deep-UV Surface-Enhanced Resonance Raman Spectroscopy. *Nano Lett.* **2016**, *16*, 7968–7973.

(61) Leijtens, T.; Eperon, G. E.; Pathak, S.; Abate, A.; Lee, M. M.; Snaith, H. J. Overcoming Ultraviolet Light Instability of Sensitized TiO 2 with Meso-Superstructured Organometal Tri-Halide Perovskite Solar Cells. *Nat. Commun.* **2013**, *4*, 1–8.

(62) Zhang, G.; Li, W.; Chu, B.; Su, Z.; Yang, D.; Yan, F.; Chen, Y.; Zhang, D.; Han, L.; Wang, J.; Liu, H.; Che, G.; Zhang, Z.; Hu, Z. Highly Efficient Photovoltaic Diode Based Organic Ultraviolet Photodetector and the Strong Electroluminescence Resulting from Pure Exciplex Emission. *Org. Electron.* **2009**, *10*, 352–356.

(63) Prosa, M.; Tessarolo, M.; Bolognesi, M.; Margeat, O.; Gedefaw, D.; Gaceur, M.; Videlot-Ackermann, C.; Andersson, M. R.; Muccini, M.; Seri, M.; Ackermann, J. Enhanced Ultraviolet Stability of Air-Processed Polymer Solar Cells by Al Doping of the ZnO Interlayer. *ACS Appl. Mater. Interfaces* **2016**, *8*, 1635–1643.

(64) Jiao, X.; Peterson, E. M.; Harris, J. M.; Blair, S. UV Fluorescence Lifetime Modification by Aluminum Nanoapertures. *ACS Photonics* **2014**, *1*, 1270–1277.

(65) Jiao, X.; Wang, Y.; Blair, S. UV Fluorescence Enhancement by Al and Mg Nanoapertures. *J. Phys. Appl. Phys.* **2015**, *48*, 184007.

(66) Pellegrotti, J. V.; Acuna, G. P.; Puchkova, A.; Holzmeister, P.; Gietl, A.; Lalkens, B.; Stefani, F. D.; Tinnefeld, P. Controlled Reduction of Photobleaching in DNA Origami–Gold Nanoparticle Hybrids. *Nano Lett.* **2014**, *14*, 2831–2836.

(67) Jha, S. K.; Mojarad, N.; Agio, M.; Löffler, J. F.; Ekinci, Y. Enhancement of the Intrinsic Fluorescence of Adenine Using Aluminum Nanoparticle Arrays. *Opt. Express* **2015**, *23*, 24719–24729.

(68) Ringemann, C.; Schönle, A.; Giske, A.; Middendorff, C. von; Hell, S. W.; Eggeling, C. Enhancing Fluorescence Brightness: Effect of Reverse Intersystem Crossing Studied by Fluorescence Fluctuation Spectroscopy. *ChemPhysChem* **2008**, *9*, 612–624.

(69) Kohl, T.; Schwille, P. Fluorescence Correlation Spectroscopy with Autofluorescent Proteins. In *Microscopy Techniques: -/-*; Rietdorf, J., Ed.; Advances in Biochemical Engineering; Springer Berlin Heidelberg: Berlin, Heidelberg, 2005; pp 107–142.

(70) Widengren, J.; Mets, Ü. Conceptual Basis of Fluorescence Correlation Spectroscopy and Related Techniques as Tools in Bioscience. In *Single Molecule Detection in Solution*; John Wiley & Sons, Ltd, 2003; pp 69–120.

# Supporting Information for

# Ultraviolet Photostability Improvement for

# Autofluorescence Correlation Spectroscopy on Label-Free Proteins

Aleksandr Barulin, Jérôme Wenger*

*Aix Marseille Univ, CNRS, Centrale Marseille, Institut Fresnel, 13013 Marseille, France*

* Corresponding author: jerome.wenger@fresnel.fr

**Contents:**

## S1. Autofluorescence brightness as a function of the number of tryptophan residues

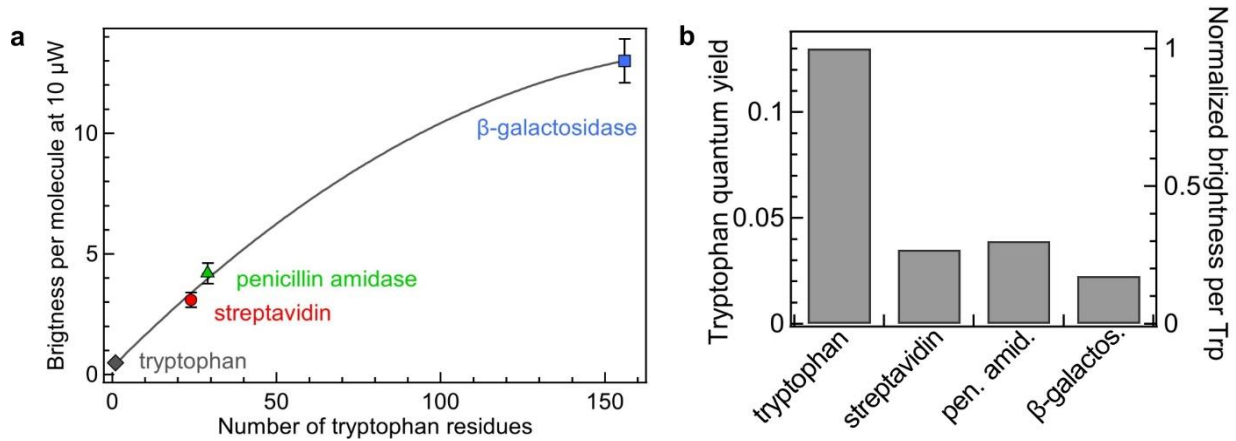

**Figure S1.** (a) Brightness per molecule (in detected photons per second) at 10 µW excitation power at 266 nm for the different samples as a function of the number of tryptophan residues per protein. The excitation power is well below the occurrence of saturation hence, antifading agents play a minimal role here and pure Hepes buffer is used to record this data. The line is a guide to the eyes. Intramolecular fluorescence quenching occurring between aminoacid residues inside the protein prevent the brightness from following a simple linear relationship with the number of tryptophan emitters. (b) Quantum yield and brightness per tryptophan residue for the different protein samples, estimated at excitation powers < 10 µW well below saturation. Here we assume a quantum yield of 13% for pure tryptophan in PBS solution and use it as a reference.

## S2. Influence of antifading agents in the absence of oxygen scavenger

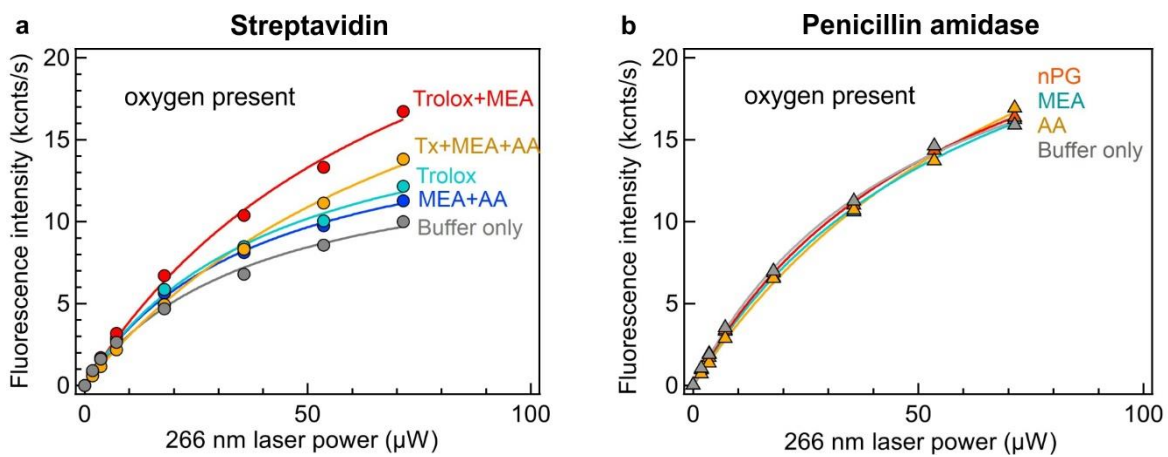

**Figure S2.** Influence of antifading agents on the autofluorescence intensity, in presence of the oxygen naturally dissolved in the solution (no oxygen scavenger is used here). The autofluorescence intensity is recorded for streptavidin (a) and penicillin amidase (b) at 1 µM concentration in the presence of different agents.

## S3. Time traces of protein autofluorescence used for FCS

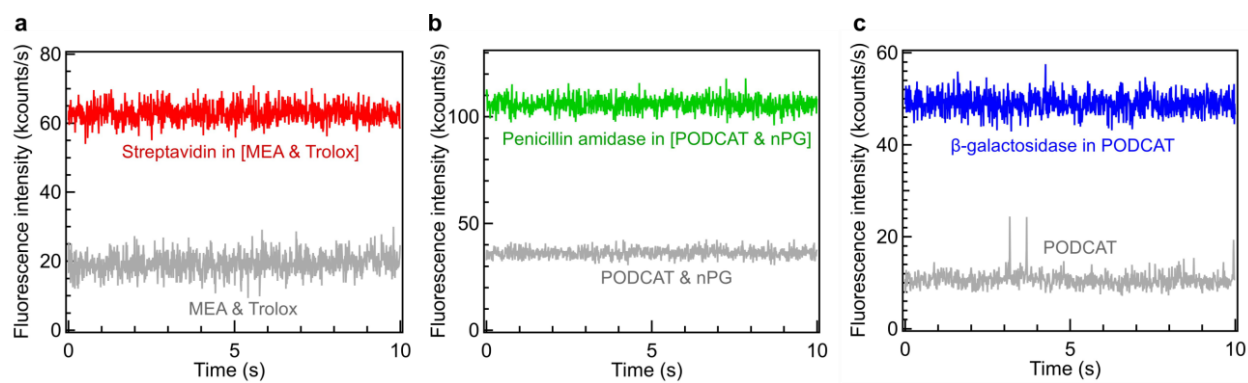

**Figure S3.** Fluorescence intensity time traces $F_{PMT}$ for the different protein samples corresponding to the FCS data in Fig. 3. The lower time traces (displayed in gray) correspond to the fluorescence background B from the buffer alone (oxygen scavengers, antifades). The concentration of protein samples is 500 nM, the average laser power is 300 µW for (a) and (b), and 150 µW for (c).

## S4. Comparison of FCS data recorded with and without antifading agent

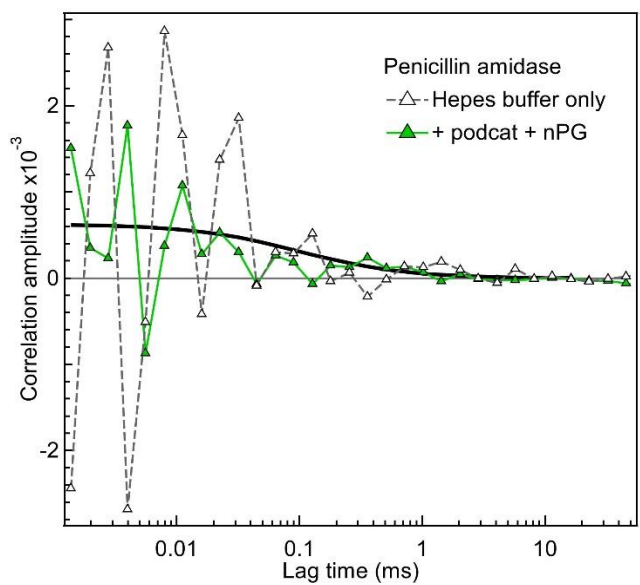

**Figure S4.** UV FCS correlation function recorded in penicillin amidase at 500 nM concentration respectively in the absence or presence of PODCAT (1.5 µM pyranose oxidase, 0.83 µM catalase, 10 wt% glucose) and 100 µM nPG. The improved fluorescence brightness brought by PODCAT and nPG significantly reduces the noise in the FCS data.

## S5. Correlation of background fluorescence from the buffer alone

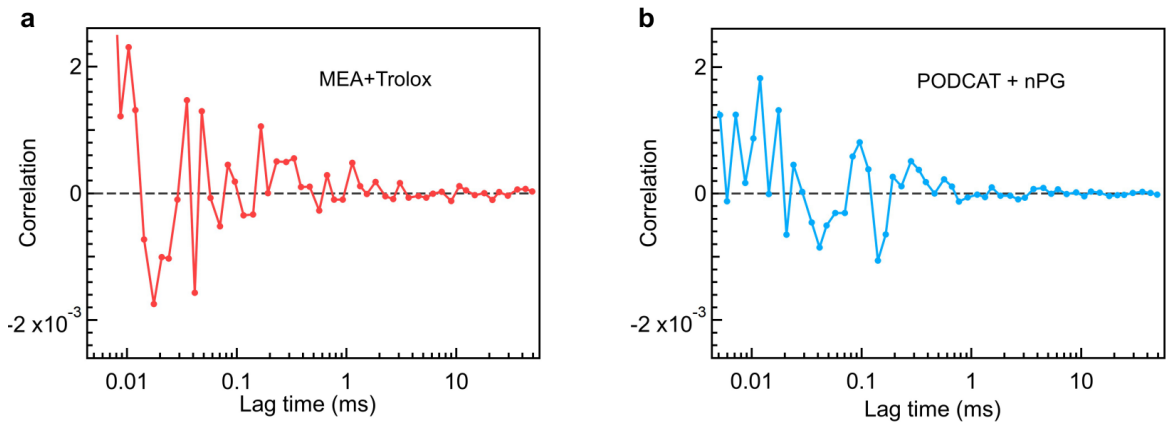

**Figure S5.** Fluorescence autocorrelation functions for the background noise B recorded on two different buffers alone: MEA 10 mM + Trolox 1 mM (a) and PODCAT + nPG 100 µM (b), in the same conditions as Fig. 3. In the absence of the protein sample, no correlation is obtained, confirming that the FCS data on Fig. 3 stems from the protein.

## S6. FCS data on dilutions series of β-galactosidase samples

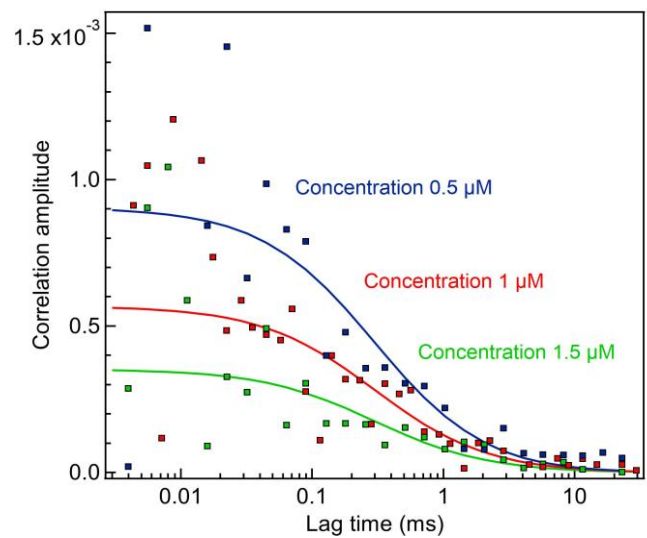

**Figure S6.** FCS correlation functions recorded at 3 different β-galactosidase concentrations. The FCS amplitude scales with the inverse of the number of molecules. Therefore, the correlation amplitude increases as the protein concentration goes down. A linear relationship between the concentration and the number of molecules is recovered, see Fig. 3e of the main document.

## S7. Increasing the viscosity on streptavidin UV FCS experiments

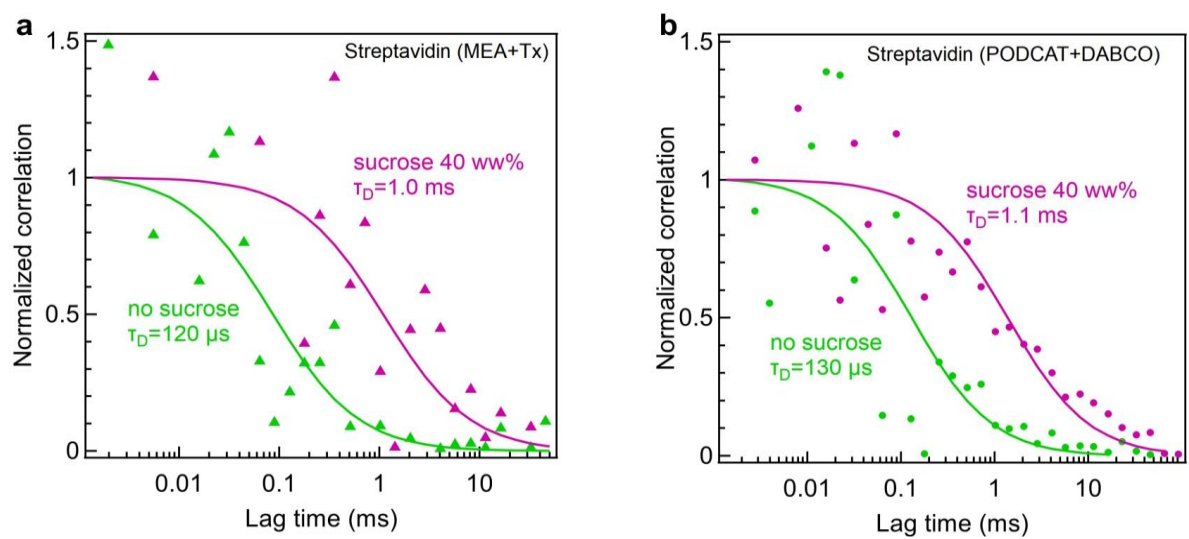

**Figure S7.** FCS correlation function for streptavidin without and with 40% w/w sucrose to increase the viscosity. Two different conditions are tested, leading to similar results: (a) streptavidin with 4 mM MEA and 1 mM Trolox, (b) streptavidin with PODCAT (1.5 µM pyranose oxidase, 0.83 µM catalase, 10 wt% glucose) and 10 mM Dabco.

## S8. Protein autofluorescence lifetime analysis

The fitting of lifetime decays is performed by Levenberg-Marquard optimization using a commercial software (Picoquant SymPhoTime 64). The program carries out an iterative reconvolution fit taking into account a contribution from the instrument response function (IRF). The region of interest in the temporal decays are set to ensure that more than 97% of all detected photons are considered. In the fitting procedure, we include non-single exponential fit due to the nature of tryptophan photochemistry. We use 3 exponential components for lifetime decay fitting, one of the components is kept fixed at 0.01 ns to interpolate the main peak. The other two components are free in the numerical fit. The intensity averaged lifetime is taken to compare the lifetimes of L-tryptophan, β-galactosidase, penicillin amidase and streptavidin.

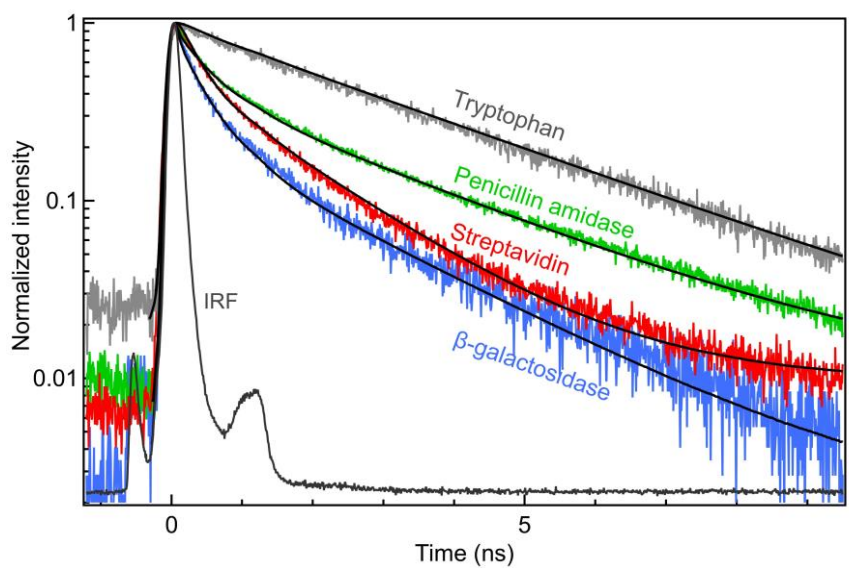

**Figure S8.** Normalized TCSPC lifetime decays for the different protein samples. No clear influence of the antifading agents or oxygen scavengers could be seen on the TCSPC decay trace. The fit results are detailed in Table S1.

**Table S1:** Parameters for the triexponential fits of the TCSPC histograms for tryptophan and proteins. $\tau_i$ indicate the lifetimes, $I_i$ are the relative intensities of each exponential component and $<\tau_{int}>$ is the intensity-averaged lifetime. The first component is always fixed at 10 ps to account for the residual backscattering of the laser light and the Raman scattered light.

| Sample | $\tau_1$ (ns) | $\tau_2$ (ns) | $\tau_3$ (ns) | $I_1$ | $I_2$ | $I_3$ | $<\tau_{int}>$ (ns) |
|---|---|---|---|---|---|---|---|
| Tryptophan | 0.01 | 0.54 | 3.07 | 0.02 | 0.02 | 0.96 | 3 |
| Penicillin amidase | 0.01 | 0.47 | 2.68 | 0.07 | 0.2 | 0.73 | 2.1 |
| β-galactosidase | 0.01 | 0.4 | 2.11 | 0.15 | 0.31 | 0.54 | 1.25 |
| Streptavidin | 0.01 | 0.3 | 1.59 | 0 | 0.27 | 0.73 | 1.24 |